\pdfoutput=1
\documentclass[12pt]{article}
\usepackage[margin=1in]{geometry}
\usepackage{amsmath}
\usepackage{titling}
\usepackage{blindtext}
\usepackage{pgfplots}
\usepackage{natbib}
\def\sin{\mathop{\rm sin}\nolimits}
\usepackage{comment}
\usepackage{setspace}
\usepackage{authblk}
\usepackage[percent]{overpic}
\usepackage[section]{placeins}
\usepackage{pbox}
\usepackage{comment}
\usepackage{titling}
\usepackage{natbib}
\usepackage{soul}
\usepackage{siunitx}
\usepackage{textcomp}
\newcommand{\GG}[1]{}

\usepackage{ragged2e}
\justifying

\begin{document}

\title{A Modified Spheromak Model Suitable for Coronal Mass Ejection Simulations}

\author[1]{Talwinder Singh}
\affil[1]{Department of Space Science, The University of Alabama in Huntsville, AL 35805, USA}

\author[2]{Mehmet S. Yalim}
\affil[2]{Center for Space Plasma and Aeronomic Research, The University of Alabama in Huntsville, AL 35805, USA}

\author[1,2]{Nikolai V. Pogorelov}

\author[3]{Nat Gopalswamy}
\affil{NASA/Goddard Space Flight Center, Greenbelt, MD 20771, USA}

\setcounter{Maxaffil}{0}
\renewcommand\Affilfont{\itshape\small}
\date{}  

\begin{titlingpage}
    \maketitle

\begin{abstract}

Coronal Mass Ejections (CMEs) are one of the primary drivers of extreme space weather. They are large eruptions of mass and magnetic field from the solar corona and can travel the distance between Sun and Earth in half a day to a few days. Predictions of CMEs at 1 Astronomical Unit (AU), in terms of both its arrival time and magnetic field configuration, are very important for predicting space weather. Magnetohydrodynamic (MHD) modeling of CMEs, using flux-rope-based models is a promising tool for achieving this goal. In this study, we present one such model for CME simulations, based on spheromak magnetic field configuration. We have modified the spheromak solution to allow for independent input of poloidal and toroidal fluxes. The motivation for this is a possibility to estimate these  fluxes from solar magnetograms and extreme ultraviolet (EUV) data from a number of different approaches. We estimate the poloidal flux of CME using post eruption arcades (PEAs) and toroidal flux from the coronal dimming. In this modified spheromak, we also have an option to control the helicity sign of flux ropes, which can be derived from the solar disk magnetograms using the magnetic tongue approach. We demonstate the applicability of this model by simulating the 12 July 2012 CME in the solar corona.     

\end{abstract}
\end{titlingpage}

\section{Introduction}\label{introduction}

Coronal Mass Ejections (CMEs) are one of the most violent events in our solar system. The total energy released in these events can range between $10^{22}$ to $10^{25}$ Joules \citep{Vourlidas02}. A CME can be ejected with speeds ranging between 20 km/s and 3500 km/s. CMEs have an average speed of 300 km/s during solar minima and 500 km/s during solar maxima \citep{Yashiro04}. Thomson scattering of white light from CME electrons allows us to calculate the total mass of CMEs, which has been found to be between $10^{14}$ and $4\times 10^{16}$ g, with an average of $10^{15}$ g \citep{Gopal10}. One of the important CME features is their magnetic flux rope structure, which is primarily responsible for the CME's geoeffectiveness. Particularly, a CME is more geoeffective if its flux rope has a large negative Bz component of magnetic field at Earth. This is because of the favorable conditions for magnetic reconnection between the flux-rope field and Earth's magnetic field in the day-side magnetosphere. 

CME predictions at 1 AU remain an area of active research. MHD simulations are clearly of importance for achieving better accuracy as compared with the simple empirical models \citep[e.g.][]{Vandas96,Brueckner98,Gopalswamy01,Gopalswamy05,Wang02,Manoharan04}. Many case studies have been done using MHD models that show reasonable agreement between the simulated and observed properties of CMEs ~\citep[][and references therein]{Manchester04, Jin17, Singh18, Singh19}.  

Current MHD CME models are broadly divided into two categories: (1) over-pressured plasmoid models, such as the blob model~\citep[e.g., see,][]{Chane05,OP99a} and (2) flux-rope-based models, such as the Titov--Demoulin model \citep{TD99}, the Gibson--Low (GL) \citep{GL98} model, and their variations. Since the magnetic flux rope of a CME is primarily responsible for its geoeffectiveness, flux-rope-based models are clearly more realistic and promising for space weather predictions.

Accurate predictions at 1 AU are impossible without constraining a CME model with observations. The parameters that must be constrained during a CME eruption are: 
1) CME speed,
2) direction,
3) orientation (the tilt angle between the horizontal and the axis of the flux rope at its apex),
4) poloidal flux,
5) toroidal flux,
6) helicity sign, and
7) mass.
In this study, we have modified the force balanced spheromak solution, so that the above-mentioned CME properties can be constrained substantially in simulations. We also discuss some of the existing methods being used to derive these parameters from the observations. In Sec. \ref{models}, we present the data-driven solar wind model and the modified spheromak model. Section \ref{data} describes various data and methods being used to find CME parameters. In Sec. \ref{results}, we present an example of the application of our simulation model. Our conclusions are presented in Sec. \ref{Conclusions}. 

\section{Simulation models}\label{models}
In this study, we perform the CME simulations in two steps. First, we create a solar wind background from 1.03 $R_\odot$ to 30 $R_\odot$ using solar synoptic magnetograms. Then the flux rope model is inserted into the domain and it erupts as a CME due to pressure imbalance. We use Multi Scale Fluid Kinetic Simulation Suite (MS-FLUKSS, \citet{Pogorelov14, Pogorelov17}). MS-FLUKSS is a highly parallelized code that can be used for MHD treatment of plasma and fluid or kinetic treatment of neutral hydrogen atoms. Both the solar wind model and the flux rope model are described in the following subsections.

\subsection{Solar wind model}\label{models1}
In this study we use the global MHD solar corona model \citep{Yalim17} driven by radial synoptic maps from the Solar Dynamics Observatory's Helioseismic and Magnetic Imager (SDO-HMI). The ideal MHD equations are solved in the solar co-rotating frame, with volumetric heating terms providing the required acceleration to the solar wind \citep{Nakamizo09}. These volumetric heating terms consist of an exponential heating function that takes the expansion factor into account and a Spitzer-type thermal conduction term that conducts heat along the magnetic field lines. The initial distribution of the magnetic field is found using the Potential Field Source Surface (PFSS) model \citep{Toth11}. The rest of the initial plasma parameters are computed using Parker's 1D isothermal solar wind solution \citep{Parker58}. We describe the implementation of boundary conditions in more detail in Sec. \ref{results}.  

\subsection{Modified Spheromak model}\label{models2}
Here, we describe a flux rope model based on the spheromak solution in which magnetic forces are balanced by plasma pressure gradient forces to create a spherical spheromak in equilibrium \citep{GL98, Lites95}. The $\nabla \cdot B = 0$ condition is specifically taken into account in this procedure. The magnetic field morphology thus achieved can be seen in Fig. 9 of \citet{Lites95}, Fig. 4 in \citet{GL98} and Fig. 1 in \citet{Singh18}. The analytical solution for the spheromak model, as given in Apendix B2 of \citet{GL98} , is:
\begin{equation}\label{Eq1}
\vec{b} = \frac{1}{r \sin \theta}\Big(\frac{1}{r}\frac{\partial A}{\partial \theta}\hat{r} - \frac{\partial A}{\partial r}\hat{\theta} + \alpha_0 A \hat{\phi}\Big),     
\end{equation}
\begin{equation}
A = \frac{4 \pi a_1}{\alpha_0^2}\Big[\frac{r_0^2}{g(\alpha_0 r_0)}g(\alpha_0 r) - r^2 \Big] \sin^2 \theta , 
\end{equation}
\begin{equation}
g(\alpha_0 r) = \frac{\sin (\alpha_0 r)}{\alpha_0 r} - \cos (\alpha_0 r) , 
\end{equation}
with $\alpha_0$ and $r_0$ related as $\alpha_0 r_0 = 5.763459$, which is the first root of the Bessel function $J_{5/2}$. Here, $r_0$ is the spheromak radius. The magnetic field strength in the spheromak is controlled by the parameter $a_1$. The plasma pressure in a spheromak is given by $P = a_1 A$. The origin of the spherical coordinate system $r$ , $\theta $ and $\phi$ is placed at the spheromak center. We can perform coordinate transformations to shift this sphere to some off-center position and also rotate it. \citet{GL98} built on this solution by including a stretching parameter that can turn a spherical torus into a tear drop shape. Several studies have been done to show the applicability of this model to simulate flux-rope-driven CMEs \citep[e.g.][]{Manchester04, Lugaz05, Jin17, Singh18}. When the force-balanced flux-rope model is superimposed with the background solar wind, the pressure imbalance inside and outside  the flux rope results in its eruption. 

There are two major drawbacks of using this model for CME simulations. Firstly, the poloidal and toroidal magnetic fluxes cannot be controlled independently in this model, since there is only one parameter $a_1$ that controls the magnetic field strength of the flux rope. For different CME sizes and magnetic strength parameters, we find that the poloidal and toroidal fluxes do not differ more than $10\%$ from each other. This is not necessarily true in an actual CME, so using the correct magnetic fluxes in the model is a key requirement for $B_z$-prediction at 1 AU. Secondly, the spheromak solution is unable to control the helicity sign of the flux rope. The helicity sign defines the direction of magnetic field line winding in the flux rope. The change in its sign can result in a completely different magnetic field at 1 AU. 

To address these shortcomings in the original spheromak solution, we propose to modify it by introducing two extra parameters $\gamma$ and $\delta$ in Eq. \ref{Eq1} so that
$$\vec{b} = \frac{1}{r \sin \theta}\Big(\gamma \frac{1}{r}\frac{\partial A}{\partial \theta}\hat{r} - \gamma \frac{\partial A}{\partial r}\hat{\theta} + \delta \alpha_0 A \hat{\phi}\Big)$$      
This new solution is no longer in the force-balance condition. This means that this model cannot be used to simulate pre-eruption, force-balanced flux ropes and their initiation phase. This model, however, can be readily used to simulate  flux ropes of erupting CMEs, which are already in a force-imbalance condition. It should be noted that erupted flux ropes of CMEs are considerably different from pre-eruptive flux ropes. This is because magnetic reconnection occurring during an eruption modifies the poloidal flux of a CME considerably \citep{Longcope07, Qiu07, Gopalswamy18}. Using such a force imbalanced flux rope model not only makes the simulation more robust, it also facilitates the input of observed magnetic flux into it. Force--imbalanced models have been successfully used in previous works as well. E.g., \citet{Manchester04} modified a force balanced Gibson-Low flux rope by reducing its density by 20\% before superimposing it on the background solar wind. They further modified density and pressure inside the flux rope such that they do not drop below 25\% of the background values. These modifications result in force imbalance in the initial flux rope itself. Similarly, \citet{Lugaz07} used a modified Titov-Demoulin flux rope by removing the strapping magnetic field lines from flux rope that were keeping it in force balanced condition. Thus, their initial flux rope was in force--imbalanced condition as well.

Our modified flux rope still satisfies the $\nabla \cdot B = 0$ condition. Plasma density inside the flux rope is assumed constant initially, and it depends on the mass of the simulated CME. Now, in this model, the toroidal flux is proportional to $a_1$ while the poloidal flux is proportional to the product $\gamma a_1$. This gives us an independent control over the two fluxes by varying $a_1$ and $\gamma$. We have explained the method of deriving the poloidal and toroidal flux of a flux rope in Appendix A. Figure \ref{gamma} shows the effect of varying $\gamma$ on the magnetic configuration of the flux rope. The poloidal flux increases proportionally to $\gamma$ while the toroidal flux remains unchanged in all cases. We show only a few magnetic field lines here to demonstrate the increase in the number of turns in response to an increasing poloidal flux. The translucent slice is colored by plasma density. Since we initially assume constant density inside the flux rope when it is added to the background solar wind, one can see the flux rope edge as an enhanced density region. In this example, the size parameter $r_0$ is 1 $R_\odot$. The flux rope is shifted by $r_1$ = 1 $R_\odot$ from the Sun center. This means that we are using roughly only one half of the spheromak in our simulation. This geometry roughly resembles a bent--tube flux rope. Therefore, the force responsible for its eruption will be the Lorentz hoop force, arising due to the curvature of the flux rope. This force is at least partially responsible for the propagation of the eruption in actual CMEs \citep{Chen17, Green18}. In this example, we have kept the flux rope in the direction S12W06 while the orientation angle was 53 degrees with respect to the solar equator. All this can be done easily by transforming the spheromak solution from a local coordinate system to a shifted and rotated coordinate system. 

\begin{figure}[!htb]
\center
\begin{tabular}{c c c}  

\includegraphics[scale=0.1,angle=0,width=5cm,keepaspectratio]{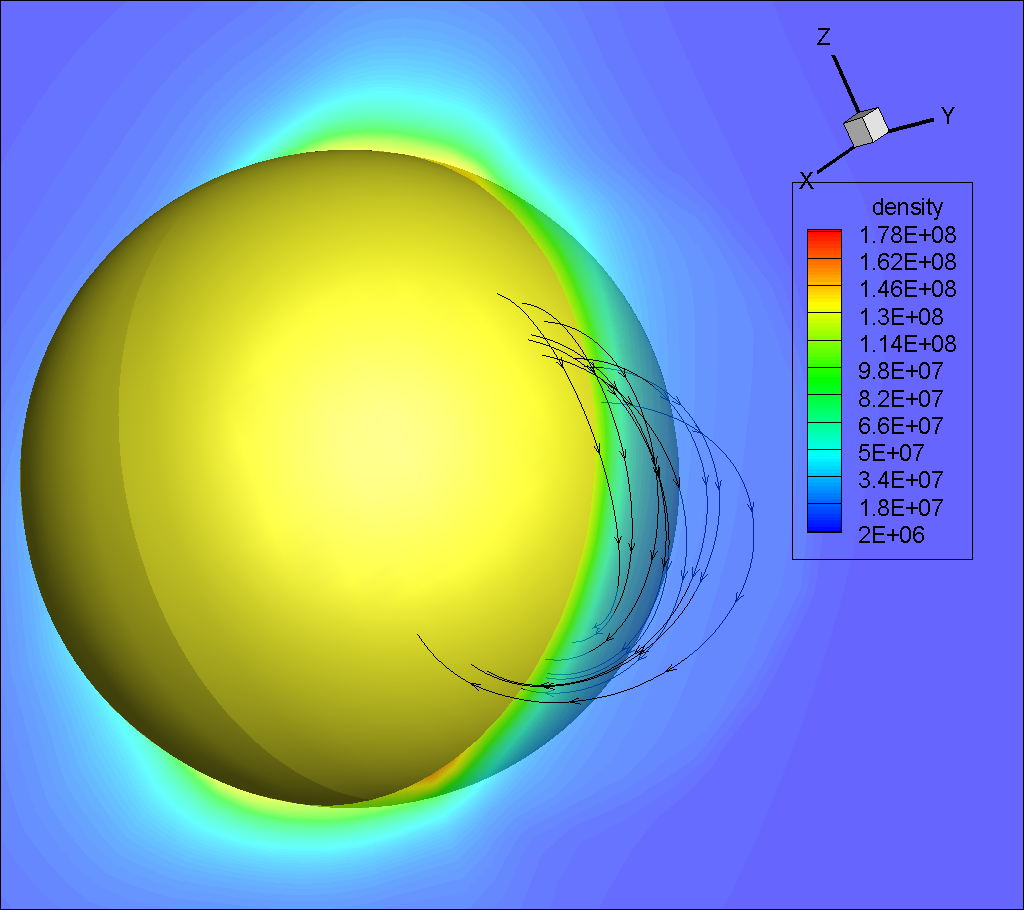}
\includegraphics[scale=0.1,angle=0,width=5cm,keepaspectratio]{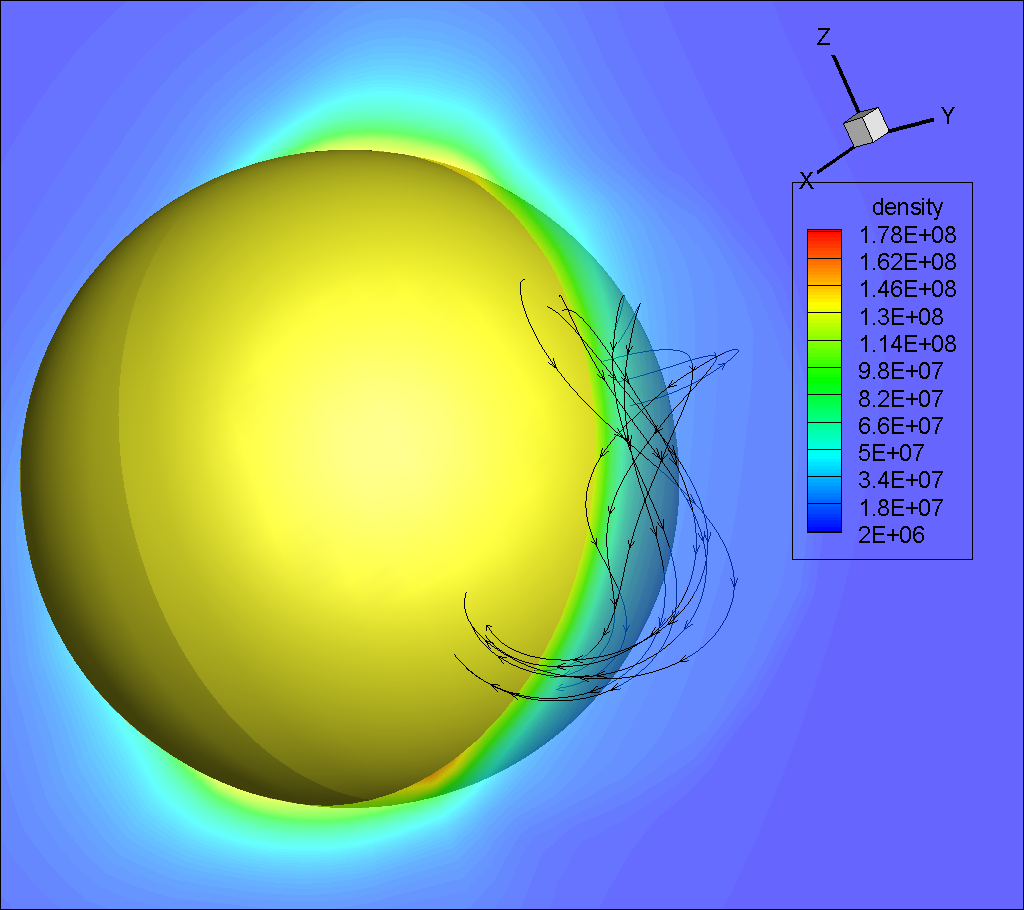} 
\includegraphics[scale=0.1,angle=0,width=5cm,keepaspectratio]{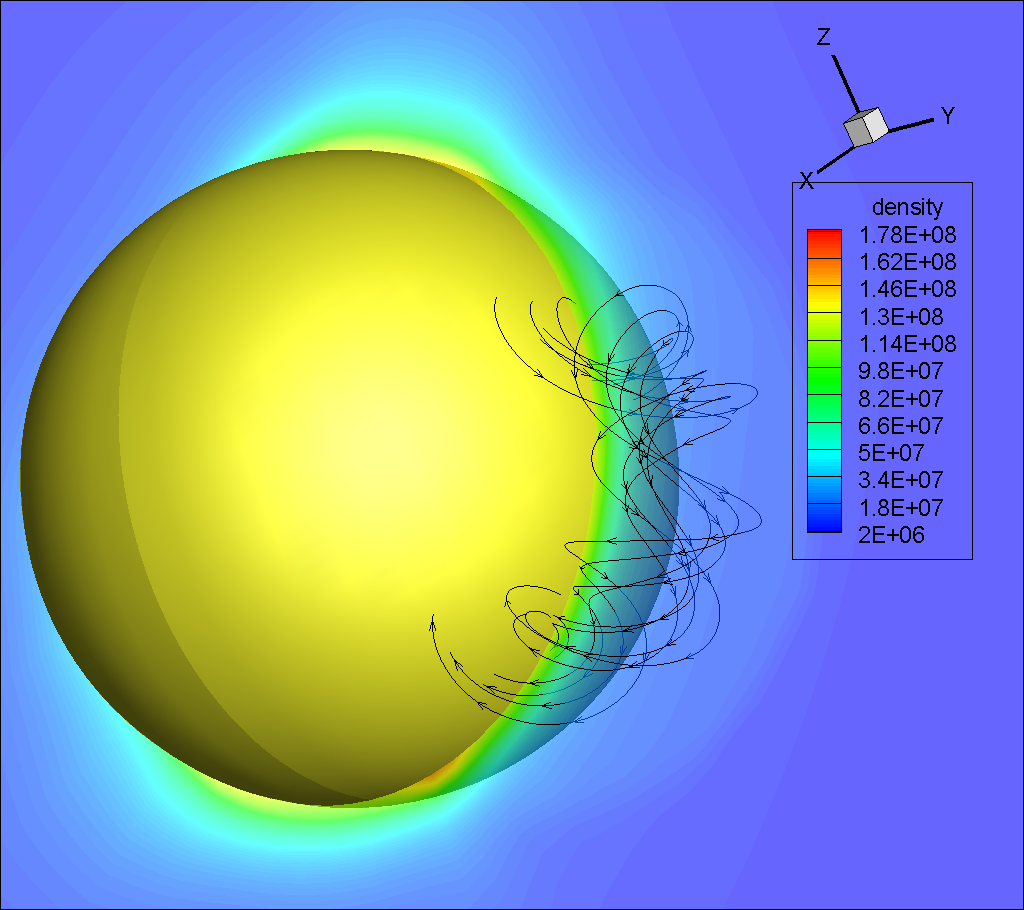} 
\end{tabular}

\caption{From \textit{left} to \textit{right}: flux rope with $\gamma$ = 1, 2, and 4 respectively. The yellow sphere represents the Sun, with magnetic field lines given by arrowed black lines. The slice through the flux rope is colored by the plasma density.}
\label{gamma}
\end{figure}

\begin{figure}[!htb]
\center
\begin{tabular}{c c}  

\includegraphics[scale=0.1,angle=0,width=6cm,keepaspectratio]{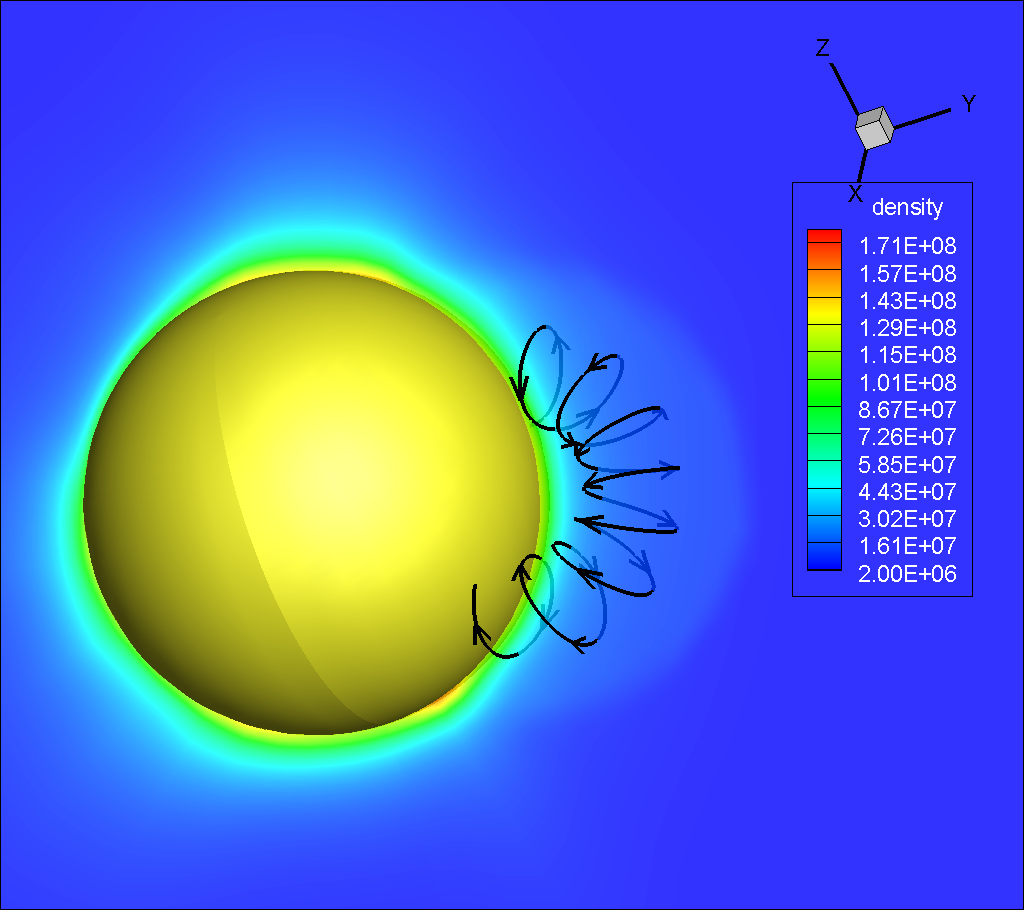}
\includegraphics[scale=0.1,angle=0,width=6cm,keepaspectratio]{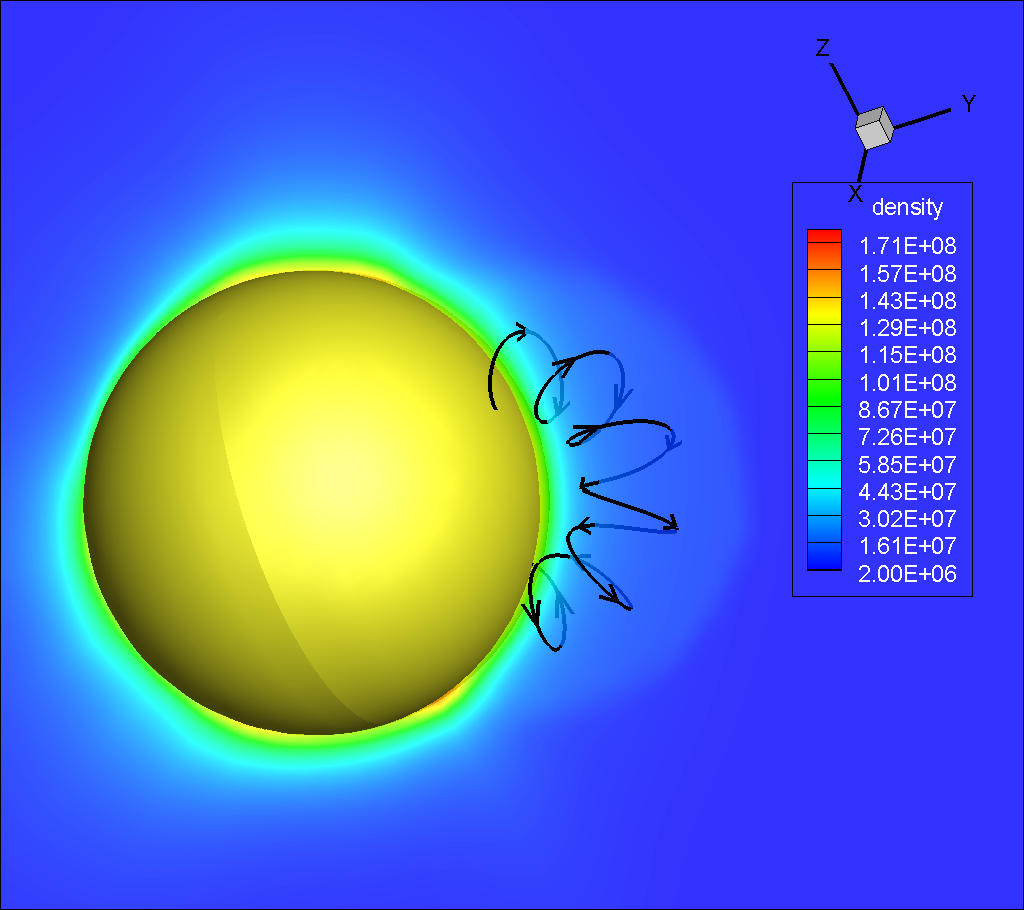} 
\end{tabular}

\caption{Flux ropes with positive (\emph{left panel}) and negative (\emph{right panel}) helicity.}
\label{helicity}
\end{figure}

The helicity sign of the flux rope can be controlled by $\delta$. The only values it can take are $\pm 1$. The plus sign will result in a positive helicity flux rope. Using $\delta = -1$ and adding 180 degrees to the orientation of the flux rope results in a negative helicity flux rope. This effect is shown in fig. \ref{helicity}. Two flux ropes are shown with the same size and magnetic field parameters. They differ only in their twist direction, i.e., the helicity sign. Here again, the flux rope edge can be inferred from the density enhancement seen in the slice.
  
The plasma $\beta$ is very low in flux ropes. Therefore, we do not include the plasma pressure into our model. The magnetic pressure dominates and is primarily responsible for CME eruption. However, we observe that introducing thermal pressure comparable to magnetic pressure in the flux rope can significantly change its eruption speed. This is equivalent to increasing the plasma energy density in the flux rope. This can be used to constrain simulated CME's speed. We can also specify the CME mass uniformly distributed throughout its volume. The stretching operation, similar to the one described by \citet{GL98}, can still be applied to this spheromak to convert the shape from sphere to a tear--drop shape. This shape conversion can bring the two legs of the flux rope closer to each other, thus reducing the width of the flux rope. This makes it possible to match the initial shape of our simulated CME with the observed CME, as shown in \citet{Singh18}.
   
\section{CME observations}\label{data}
In this section we discuss the measurable properties of CMEs, which can be used to constrain our CME model. We will briefly discuss how each quantity can be derived from observations. We use data from Solar Terrestrial Relations Observatory (STEREO) and Solar and Heliospheric Observatory (SOHO) coronagraphs, SDO magnetograms and SDO extreme ultraviolet (EUV) images to get these CME properties.

\subsection{Speed, Direction and Orientation}\label{data1} 
\begin{figure}[!htb]
\center
\includegraphics[scale=0.1,angle=0,width=15cm,keepaspectratio]{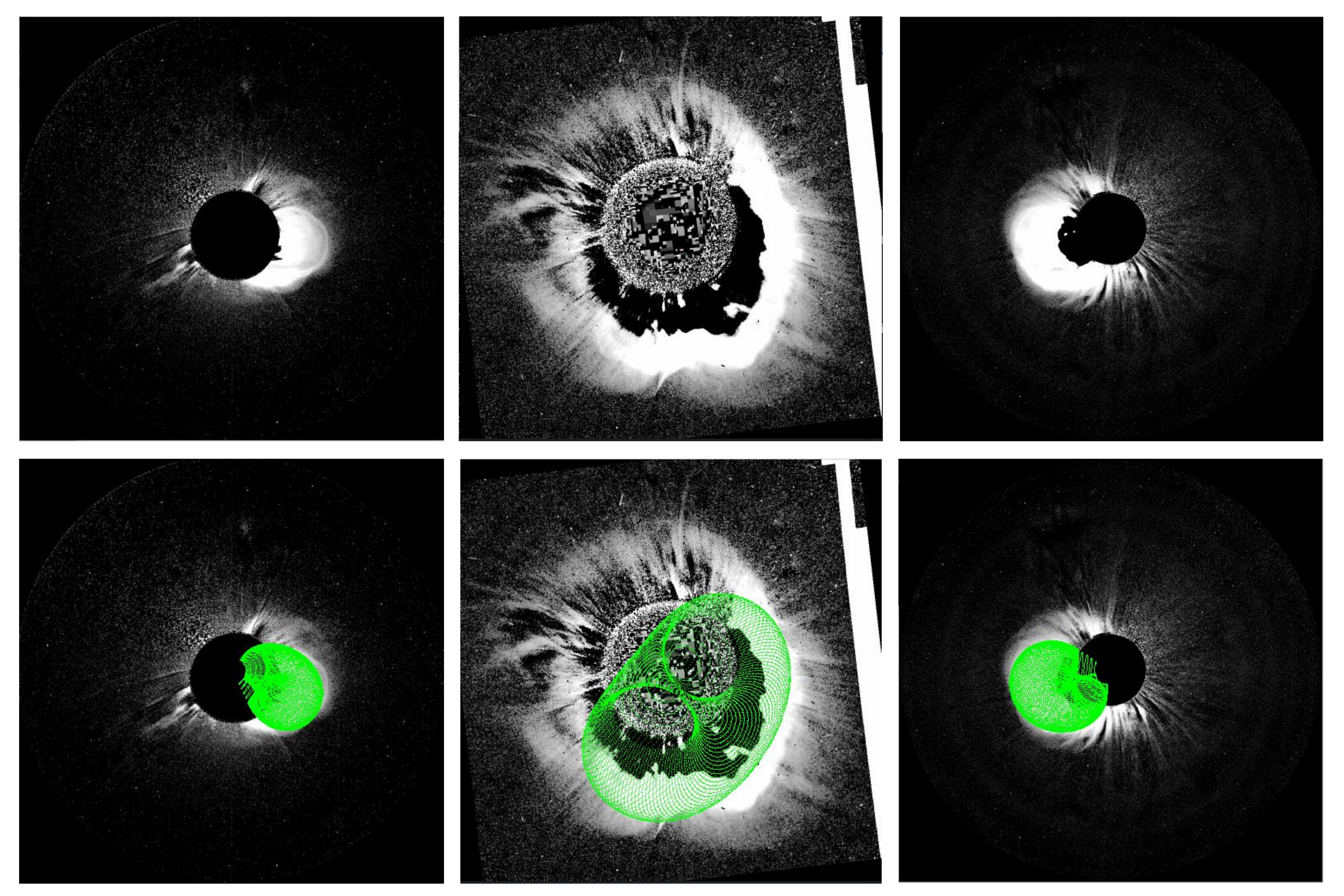}
\caption{(\textit{Top panel, from left to right}) July 12, 2012 CME seen in STEREO B Cor2, SOHO C2 and STEREO A Cor2 coronagraphs respectively. (\textit{Bottom panel}) The same images overlapped with GCS model.}
\label{GCS}
\end{figure}

Coronagraphs are one of the primary instruments used to study CME evolution in the corona. However, a single viewpoint image is not sufficient to resolve the 3-D structure of a CME. To overcome these limitations, the twin STEREO spacecraft \citep{Kaiser08} were launched to see the same CME from multiple viewpoints. The triangulation techniques can then be used to study the CME in 3D space. This helps us to remove the projection effects and get the true kinematic properties of a CME. One of the best models for utilizing the three viewpoints of STEREO A\&B and SOHO coronagraphs to find 3D CME evolution is the Graduated Cylindrical Shell (GCS) model \citep{Thernisien11}. This model fits a typical CME shape, i.e., the curved front with conical legs, to observations, giving us an estimate for the CME height, direction and orientation with respect to solar equator. Figure \ref{GCS} shows this model applied to a CME that erupted on July 12, 2012. Fitting this model for a time series can be used to estimate its speed from the height-time profile \citep[see e.g.][]{Hess14}. 

\subsection{Poloidal Flux}\label{data2}
The poloidal magnetic flux of a CME can be measured from the reconnected flux either using flare ribbons or Post Eruption Arcades (PEAs). \citet{Gopalswamy17} showed that both of these methods are highly correlated. {The PEA method requires EUV data only at the time when the PEA structure has fully matured, typically in the decay phase of the flare, whereas the flare ribbon method requires 1600 \r{A} data throughout the flaring time.} Since the PEA method is more robust and easy to implement, \citet{Gopalswamy18} propose to use it in their Flux Rope from Eruption Data (FRED) model. When the overlaying magnetic field lines gets stretched and reconnected during a flux rope eruption, one half of the lines reconnect down in the active region forming PEAs and the other half contributes to the poloidal flux of the flux rope. This process can be understood more clearly looking at the {flux rope eruption} cartoon in Fig. 1 of \citet{Klein17}. \citet{Gopalswamy17} show that the poloidal flux of the CME is half the unsigned flux in the area covered by the PEAs. 

\begin{figure}[!htb]
\center
\begin{tabular}{c c}  

\includegraphics[scale=0.1,angle=0,width=6cm,keepaspectratio]{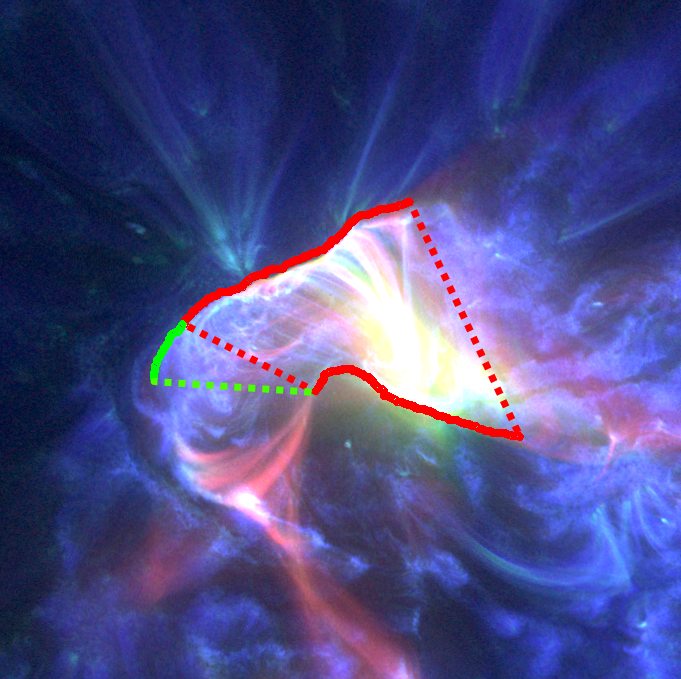}
\includegraphics[scale=0.1,angle=0,width=6cm,keepaspectratio]{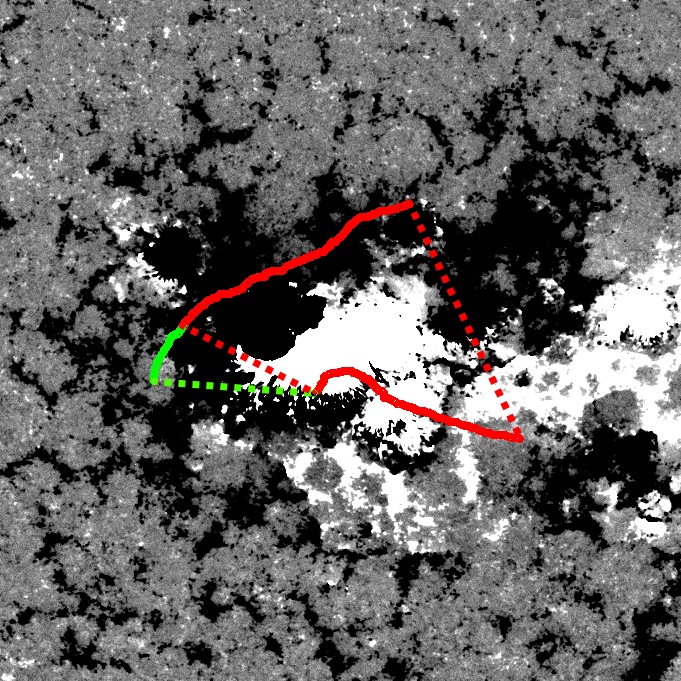} 
\end{tabular}

\caption{{(\textit{Left panel}) SDO AIA 94, 131, and 193 \r{A} composite data on 12 July 2012 22:30 UT. (\textit{Right panel}): SDO HMI magnetogram on 12 July 2012 16:10 UT. In both the cases, the area enclosed by PEAs is shown with red contours. The poloidal flux of the erupted CME is half the unsigned flux in this area. The area enclosed by the green contours has some ambiguity on whether it is a part of the PEA or some pre-eruption loops. Including this area into the PEA calculation increases poloidal flux by 9\%. The cutout size is 300 Mm $\times$ 300 Mm.}}
\label{PEA}
\end{figure}

{The left panel of Fig. \ref{PEA} shows the PEA for 12 July 2012 eruption in SDO AIA 94, 131, and 193 \r{A} composite data at 22:30 UT, when the PEA has fully matured, $\sim6$ hrs after the flare started. We chose these wavelengths because they are sensitive to temperatures exceeding 5 MK (See Fig. 1 in \citet{Cheung2015} for SDO response functions in different wavelengths). The PEAs are composed of hot loops and can be best seen in high temperature observations. We then trace the footpoints of the PEA loops,  shown here as red contours. The area between these contours and the red dotted lines should give us the area covered by the PEA. The area spanned by the green contour marks is the area of ambiguity, which can be a part of either the PEA structure or of pre-existing coronal loops. The right panel of Fig. \ref{PEA} shows the corresponding active region in SDO HMI data, along with the same contours. The poloidal flux calculated using just the area spanned by red contours is $1.35\times10^{22}$ Mx. Including the green area in our calculations increases the poloidal flux to $1.47\times10^{22}$ Mx, a 9\% increase. This gives us an estimate of the subjective error possible due to the manual PEA area selection in this method. \citet{Gopalswamy18} found the poloidal flux for the same event to be $1.42\times10^{22}$ Mx using just the 193 \r{A} data, a very similar result compared with the one we obtained using the composite image. Therefore, it may be sufficient to use just the 193 \r{A} data for this analysis.}

\subsection{Toroidal Flux}\label{data3}
The toroidal flux of a CME, also known as the axial flux, or the core flux in the literature, can be found if the footpoints of the erupting CME are found \citep[see, e.g., ][]{Webb2000}. \citet{Dissauer18} describe a way to do this using the coronal dimming. When a flux rope erupts, a clear dimming is seen in EUV images. This is due to the mass loss during the eruption. \cite{Dissauer18} show that this dimming can be separated into two categories, 1) core dimming and 2) secondary dimming using appropriate thresholds. These core dimming regions show higher mass loss and are seen during the first 30 minutes of the eruption. {Once the core dimming regions have been found, the toroidal flux of a CME is given by the unsigned average magnetic flux in the positive and negative polarity footpoints.}

\begin{figure}[!htb]
\center
\begin{tabular}{c c c}  

\includegraphics[scale=0.1,angle=0,width=5cm,keepaspectratio]{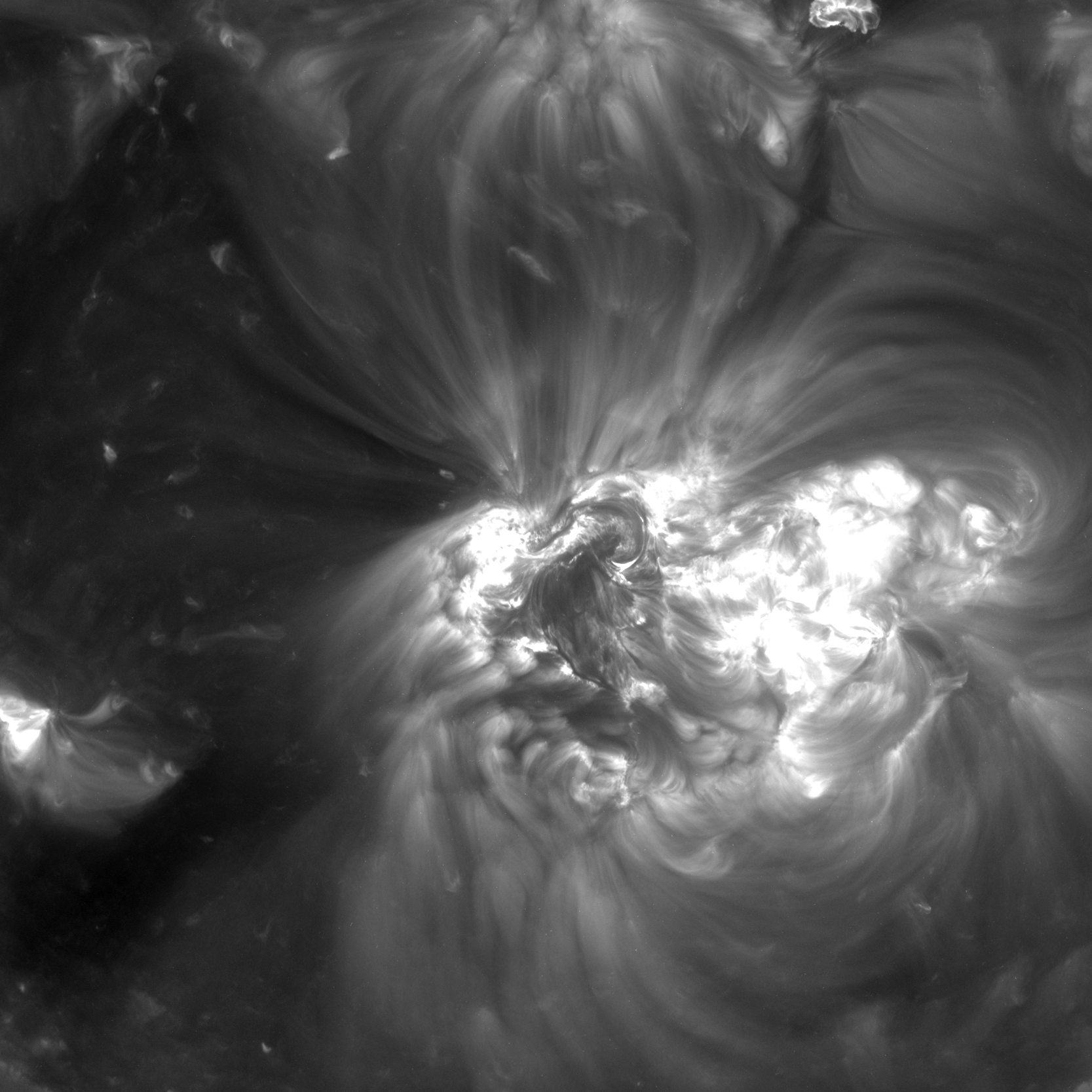}
\includegraphics[scale=0.1,angle=0,width=5cm,keepaspectratio]{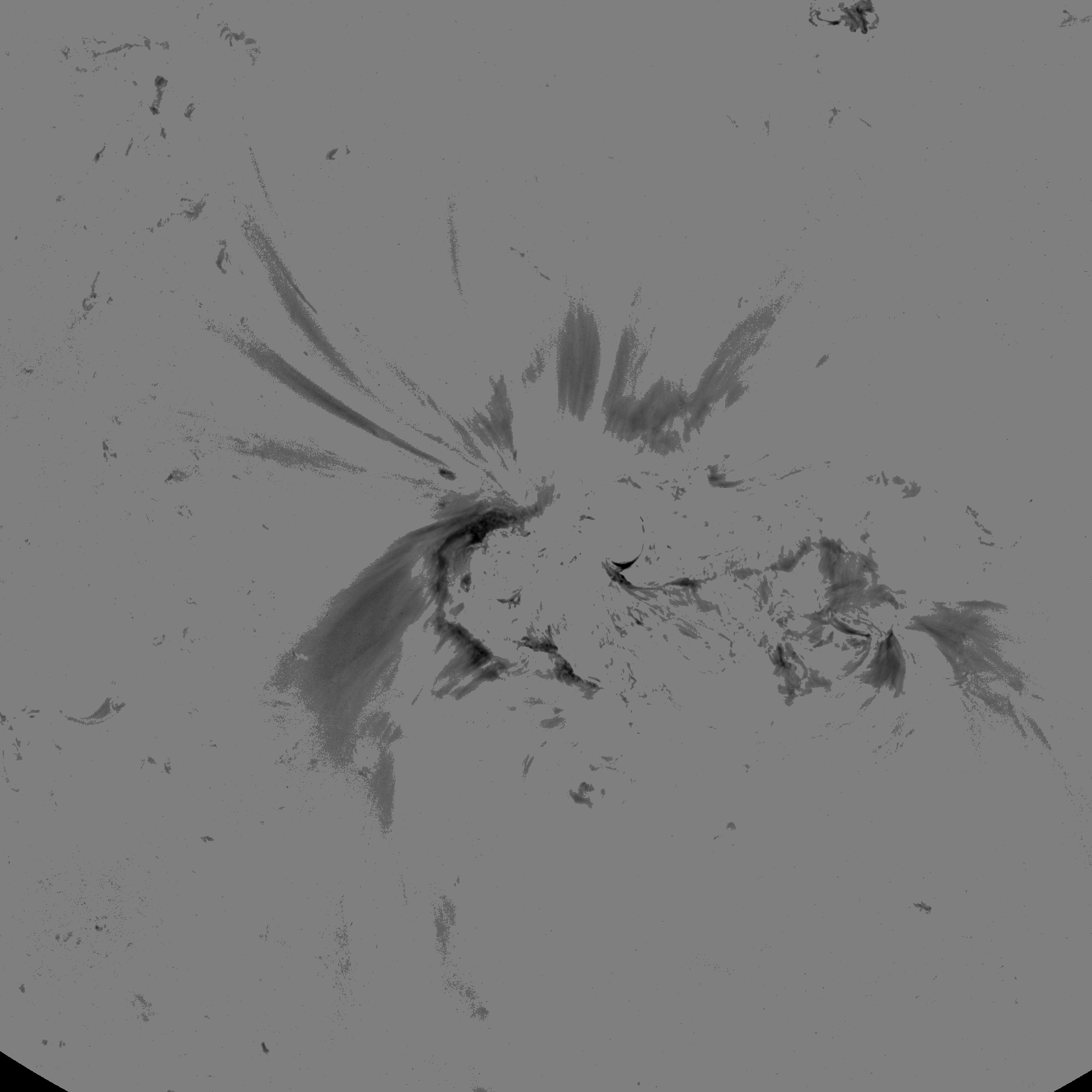} 
\includegraphics[scale=0.1,angle=0,width=5cm,keepaspectratio]{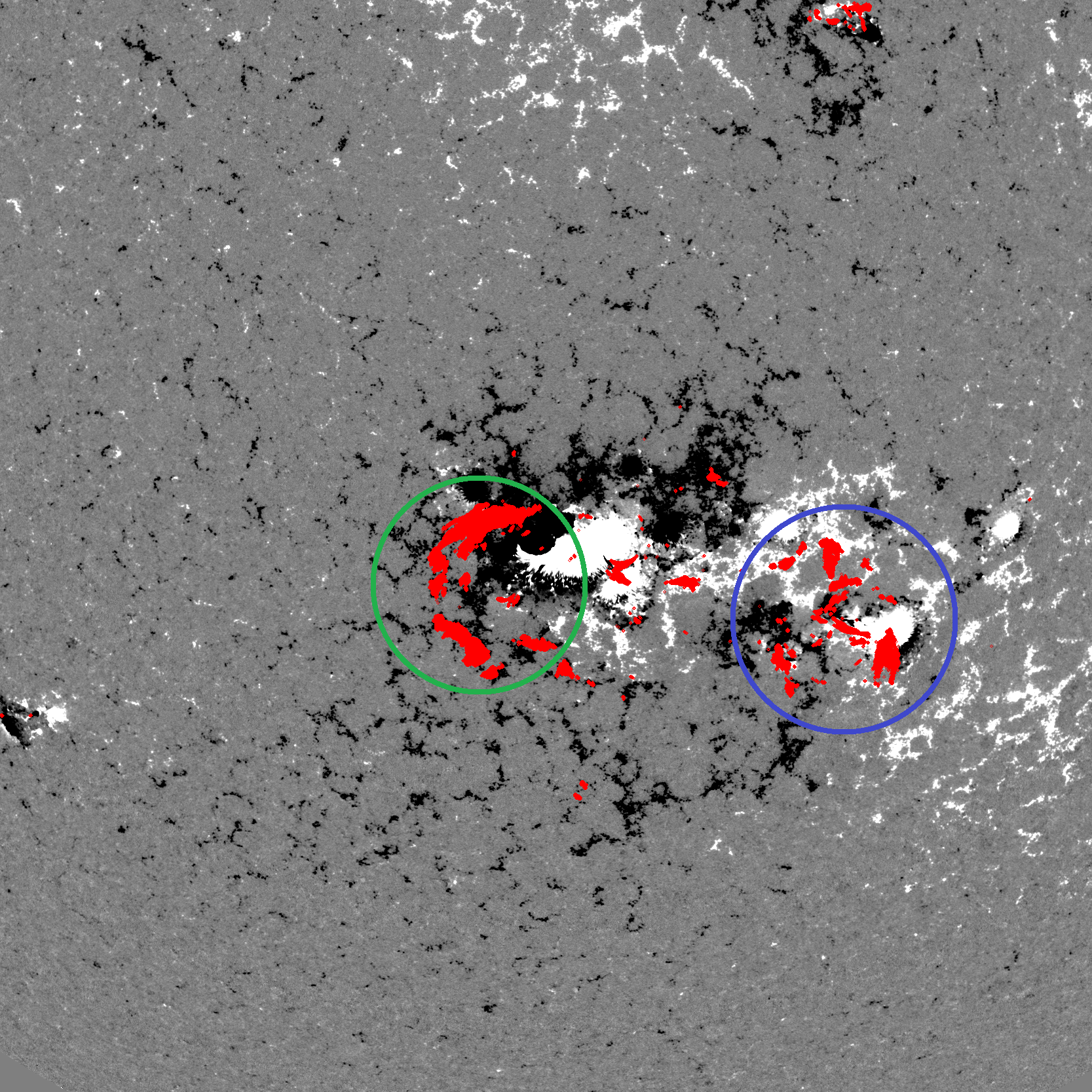} 
\end{tabular}

\caption{{(\textit{Left panel}) SDO AIA 193 \r{A} data showing the pre-eruption median image of the active region. (\textit{Middle panel}) The minimum intensity logarithmic ratio image showing coronal dimming after the eruption. (\textit{right}) The core dimming pixels are plotted over the SDO HMI magnetogram. This data is for July 12, 2012 15:40 UT. The violet and green circles enclose the areas containing the footpoints of the flux rope, predominantly in the positive and negative flux regions, respectively. The flux in these core dimming regions provides an estimate for the toroidal flux inside a CME. (Cutout size is 1000 arcsec)}}
\label{dimming}
\end{figure}

{In Fig. \ref{dimming}, we show the procedure we followed to find the toroidal flux in the 12 July 2012 CME. We use 193 \r{A} data in our analysis, which is shown to be most suitable to study coronal dimming by \cite{Dissauer18}. A pre-event image is found by taking a pixel-by-pixel median of 10 images within a half hour before the eruption starts (left panel of Fig. \ref{dimming}). Then, during 1 hour after the eruption has started, we use 2-minute-cadence data to detect the pixels where the logarithm of the ratio of any post-event image data and the pre-event image data falls below -0.19. These are the regions showing coronal dimming. We also record the lowest value of logarithmic ratios during the selected time interval among these pixels. This forms a minimum intensity logarithmic base ratio image which has non-zero values only in the coronal dimming pixels. This is shown in the middle panel of Fig. \ref{dimming}. Similarly, a minimum intensity base difference image is formed by taking differences at each pixel, rather than logarithmic ratios. Once these two minimum maps have been formed, we find a subset of the pixels that reside in the core dimming regions using the following thresholds from \citet{Dissauer18}}.
$$A=\bar{I}_{BD} - 0.6\sigma_{BD}$$
$$B=\bar{I}_{LBR} - 0.6\sigma_{LBR}$$
{Here, $\bar{I}_{BD}$ and $\bar{I}_{LBR}$ are the mean intensities over all pixels in the minimum-intensity logarithmic base ratio image and the minimum-intensity base difference image, respectively. $\sigma$ values give the corresponding standard deviations. The pixels flagged as core dimming pixels are shown in red color in the right panel of Fig. \ref{dimming}. We can see that one group (inside a violet circle) represents the positive footpoint, whereas the other (inside a green circle) represents the negative footpoint of the erupting flux rope. The unsigned flux in these core dimming pixels can be used as an estimate for the toroidal flux. The toroidal flux calculated using these pixels is $2.13\times10^{21}$ Mx. By changing the thresholds $A$ and $B$ by $\pm5\%$, for error estimation, as suggested by \citet{Dissauer18}, we find that the toroidal flux varies between $1.97\times10^{21}$ Mx and $2.31\times10^{21}$ Mx. Therefore, the toroidal flux varies by $\approx\pm8\%$. We notice that the toroidal flux for this CME is $\approx15\%$ of its poloidal flux. Such a CME cannot be reliably simulated by the original spheromak model because the poloidal and toroidal fluxes do not differ by more than 10\% from each other in this model when it is inserted near the Sun as discussed in Sec. \ref{models2}.}

\subsection{Helicity sign}\label{data4}
\begin{figure}[!htb]
\center
\includegraphics[scale=0.1,angle=0,width=10cm,keepaspectratio]{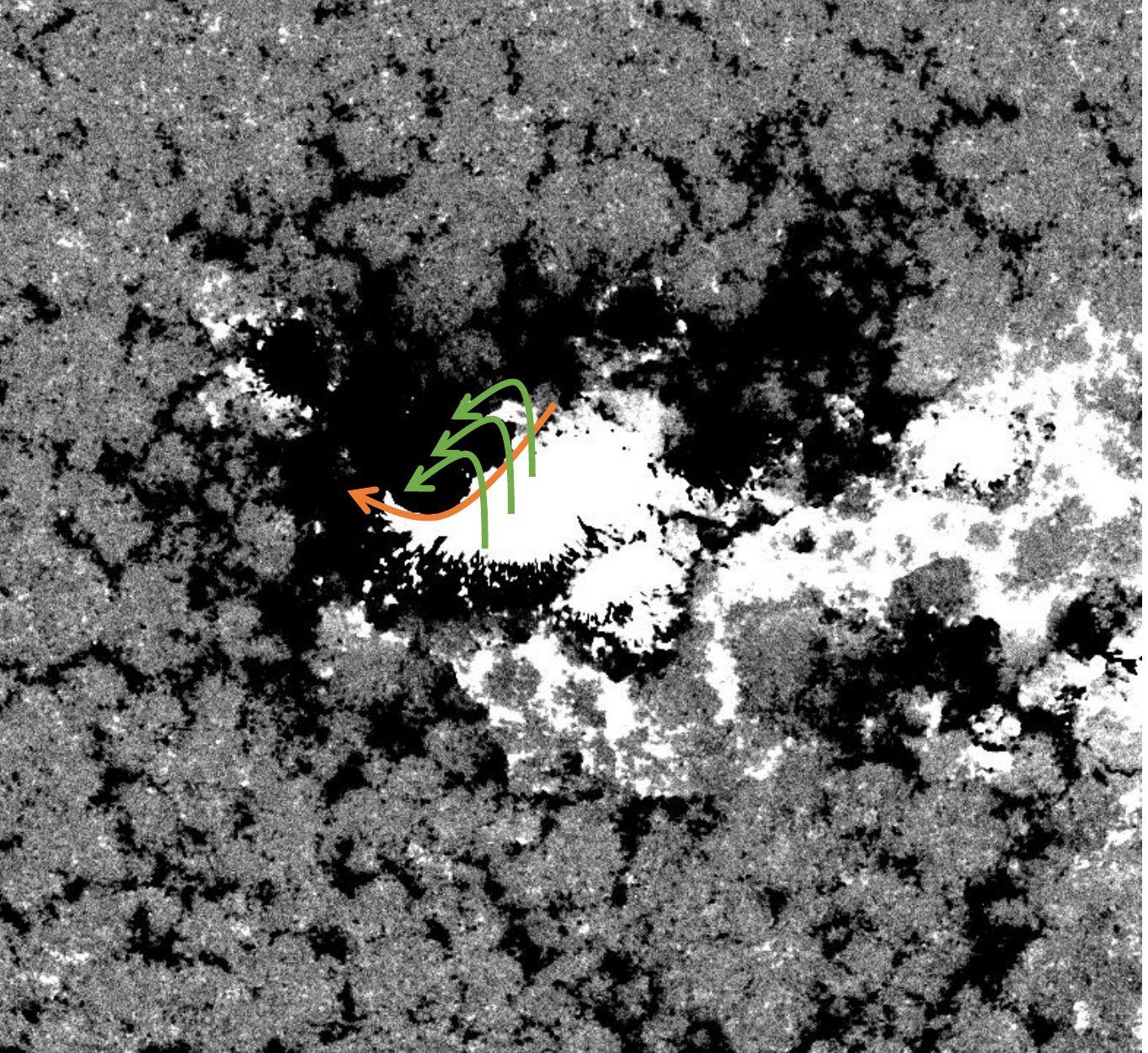}
\caption{Helicity sign estimates can be made by following the axial field lines (orange) with the thumb and curving the fingers along overlying field lines (green). This example is for 12 July 2012 CME and shows that the erupted flux rope will have positive helicity., since the field lines are traced properly by right hand.}
\label{helicity2}
\end{figure}

The helicity sign of a flux rope can be determined from the pre-eruptive active region magnetic field configuration \citep[][]{Bothmer1998}. For example, \citet{Luoni11} show how the magnetic tongues in the ARs can be used to estimate the helicity sign in the overlaying flux ropes. The helicity sign is easy to determine if one can estimate the direction of magnetic field lines on the axis of the flux rope and the overlaying loops, which eventually contribute to the poloidal flux of a CME during an eruption, as discussed in Sec. \ref{data2}. In Fig. \ref{helicity2}, we show how the helicity sign can be found if we know the flux rope footpoints and the neutral line (NL) above which the flux rope exists in corona. This NL is typically in the middle of the PEA area. The method to find footpoints has already been discussed in Sec. \ref{data3}. If we follow the NL in the direction from a positive footpoint to a negative footpoint with the thumb and curve the fingers in such a way that they follow the overlaying field, we can tell whether the helicity sign is positive or negative based on whether the right or left hand follows the lines properly. In the example shown in Fig. \ref{helicity2} for July 12, 2012 magnetogram, we know that the axial field is directed from the right to the left, since the positive footpoint is on the right as seen in Fig. \ref{dimming}. This determines the direction of orange arrow in Fig. \ref{helicity2}. The overlying field lines are shown in green color, while the direction is simply from positive to negative polarity. We can see that this configuration is consistent with the right-hand rule. Therefore, the helicity sign is positive in the erupting flux rope. The helicity sign of this event was found by \citet{Gopalswamy18} using a different method.

\subsection{Mass}\label{data5}
Mass of a CME can be found using the CME brightness in the coronagraph images. The brightness is due to the Thomson scattering of photospheric light by the plasma electrons \citep{Billings66}. By integrating over the CME area and removing the projection effects using multiple coronagraph viewpoints, we can calculate the true mass of a CME \citep{Colaninno09}.  

\section{Results}\label{results}

We will show now the applicability of our approach to an observed CME. We choose the July 12, 2012 CME as an example. This CME erupted from AR 11520 at 15:54 UT. The location of the AR was 15 degrees south and 1 degree west at the time of eruption. The reason for choosing this event is that the source active region is close to solar disk center, thus enabling more accurate estimation of the  AR magnetic field strengths. At this time, the STEREO spacecraft were observing at nearly right angles to the Sun-Earth line. This increases the accuracy of the GCS method,  thus allowing us to accurately determine the speed, direction, and orientation of the CME. Moreover, this event has been studied in detail by many authors \citep[see e.g.][]{Gopal13, Hess14, Shen14, Gopalswamy18}, thus allowing us to verify the accuracy of our derived quantities.

We determined the properties of this CME using the methods described in the previous section.
\begin{enumerate}
\item The CME speed was found to be 1210 km/s at the height of 15 $R_\odot$. This was found by fitting a quadratic function to the height-time profile found using the GCS method. A linear fit to the height-time plot gives us a speed of 1265 km/s for this CME.
\item The direction of the CME was found to be at 12 degrees south and 8 degrees west. This is a little different from the source active region location, thus the CME showed a slight deflection in the lower corona.
\item The CME flux rope orientation, found using the GCS method, was 53 degrees with respect to solar equator. These GCS results are consistent with the ones reported in \citet{Gopalswamy18}.
\item {The poloidal flux of the CME was found to be  between $1.35\times10^{22}$ Mx and $1.47\times10^{22}$ Mx using the PEA method}.
\item {The toroidal flux of the CME was found to be between $1.97\times 10^{21}$ Mx and $2.31\times 10^{21}$ Mx using the coronal dimming method.}
\item As shown in Fig. \ref{helicity2}, the CME flux rope has a positive helicity sign.
\item The mass of the CME was found to be $1.65\times 10^{16}$ g. This is the true mass of the CME with projection effects removed using the multiple viewpoints of STEREO.
\end{enumerate}

The CME simulation is carried out in two steps. First we create a solar coronal MHD background and then introduce the flux rope model in it. The coronal background is created in a fully spherical simulation domain, which extends from 1.03 $R_\odot$ to 30 $R_\odot$. The SDO HMI synoptic map for Carrington rotation 2125 is used at the inner boundary of the domain located just above the transition region at 1.03 $R_\odot$. We relax the initial PFSS magnetic field to obtain a steady state solution. We used the Total Variation Diminishing (TVD), finite volume Rusanov scheme \citep{KPS01} to compute the numerical fluxes and the forward Euler scheme for time integration. In order to satisfy the solenoidal constraint, we use the \citet{Powell99} approach. All simulations are performed in the frame corotating with the Sun. MS-FLUKSS is built on Chombo library, which ensures a highly parallelized implementation of our numerical schemes \citep{Pogorelov17}. 

At the inner boundary of the computational domain, at 1.03 $R_\odot$, we specify the differential rotation~\citep{KHH93a} and meridional flow~\citep{KHH93b} formulae for determining the horizontal velocity components at the ghost cell centers. Density and temperature are kept constant as $n=1.5 \times 10^{8} \,\textrm{cm}^{-3}$ and $T=1.3 \times 10^{6}$ K, respectively. The radial velocity component is imposed to be zero at the boundary surface. The radial magnetic field component is imposed from the magnetogram data. The transverse magnetic field components are extrapolated from the domain to the physical ghost cells below the inner radial boundary. No boundary conditions are required at the outer boundary of the domain, since it is located beyond the critical point, where the plasma flow is superfast magnetosonic.

\begin{figure}[!htb]
\center

\includegraphics[scale=0.1,angle=0,width=13cm,keepaspectratio]{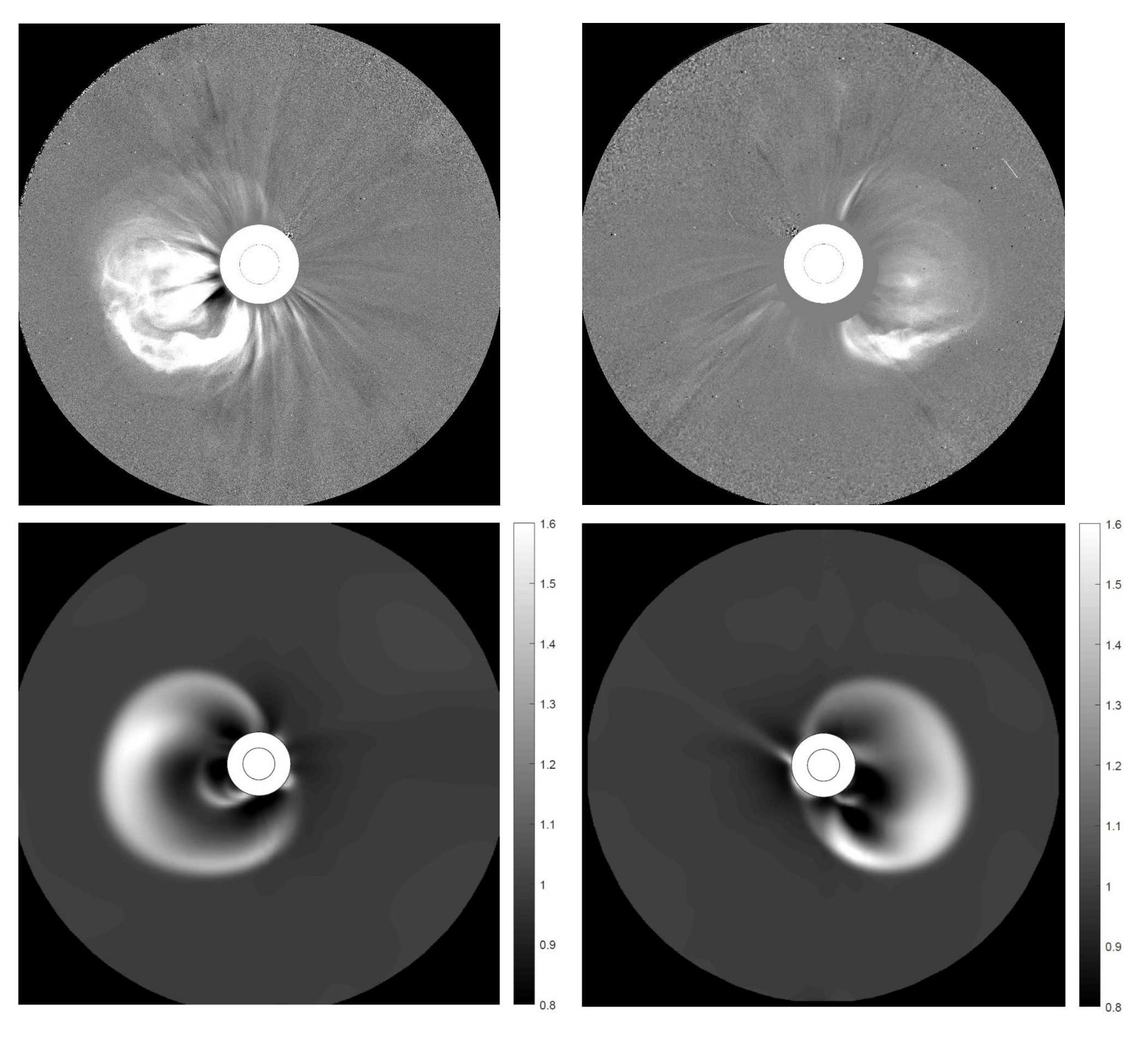}

\caption{(\textit{Top}) CME as seen by STEREO A (left) and B (right) COR 2 at 12-July-2012 17:54 UT. (\textit{Bottom}) Synthetic white light images of the simulated CME at same height, with observer fixed at the location of STEREO A (left) and B (right). These images are created by using the ratio of 
line-of-sight-integrated brightness of the post-event and pre-event images. The color map is based on this ratio.}
\label{white_light}
\end{figure}

\begin{figure}[!htb]
\center
\begin{tabular}{c c c}  

\includegraphics[scale=0.1,angle=0,width=5cm,keepaspectratio]{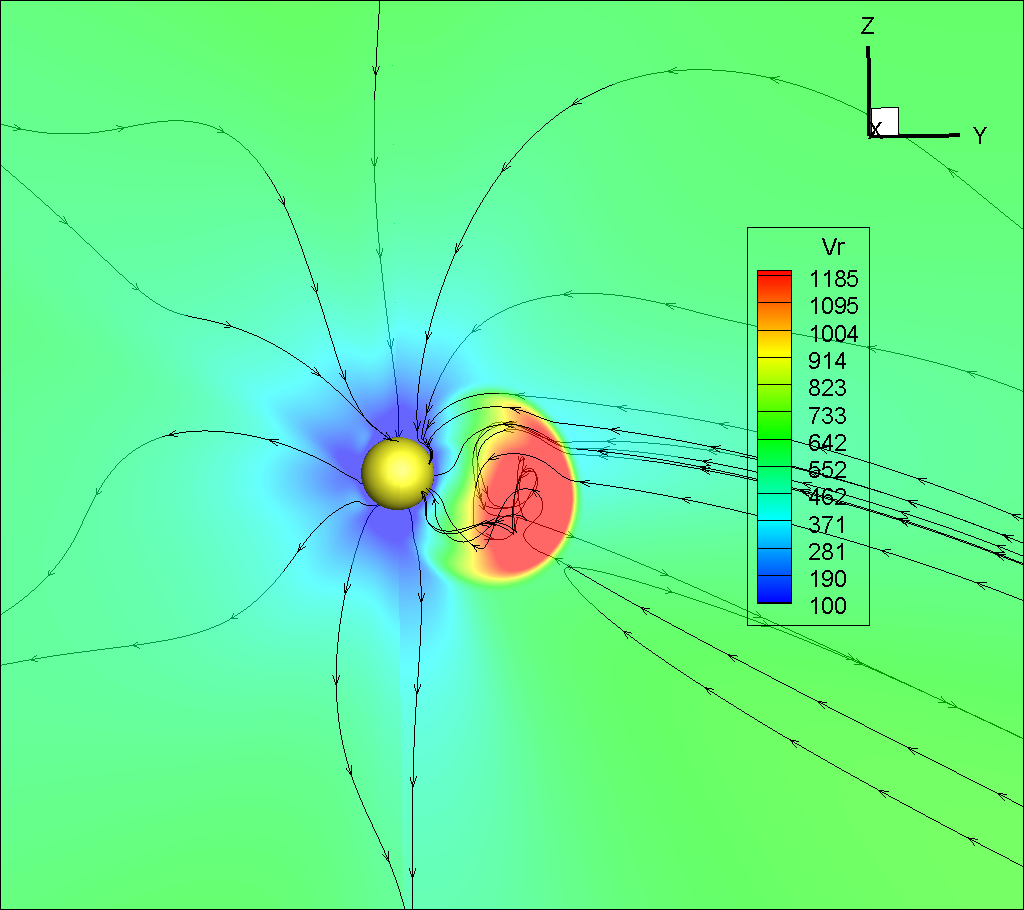}
\includegraphics[scale=0.1,angle=0,width=5cm,keepaspectratio]{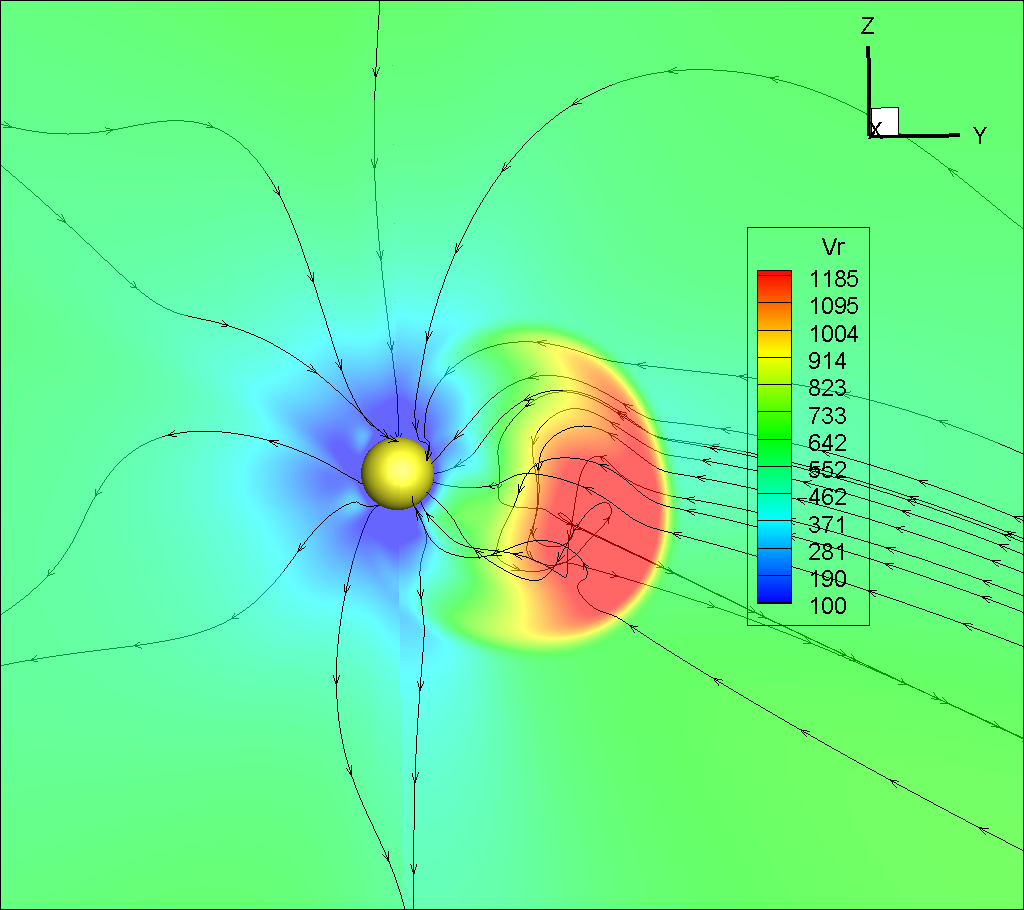} 
\includegraphics[scale=0.1,angle=0,width=5cm,keepaspectratio]{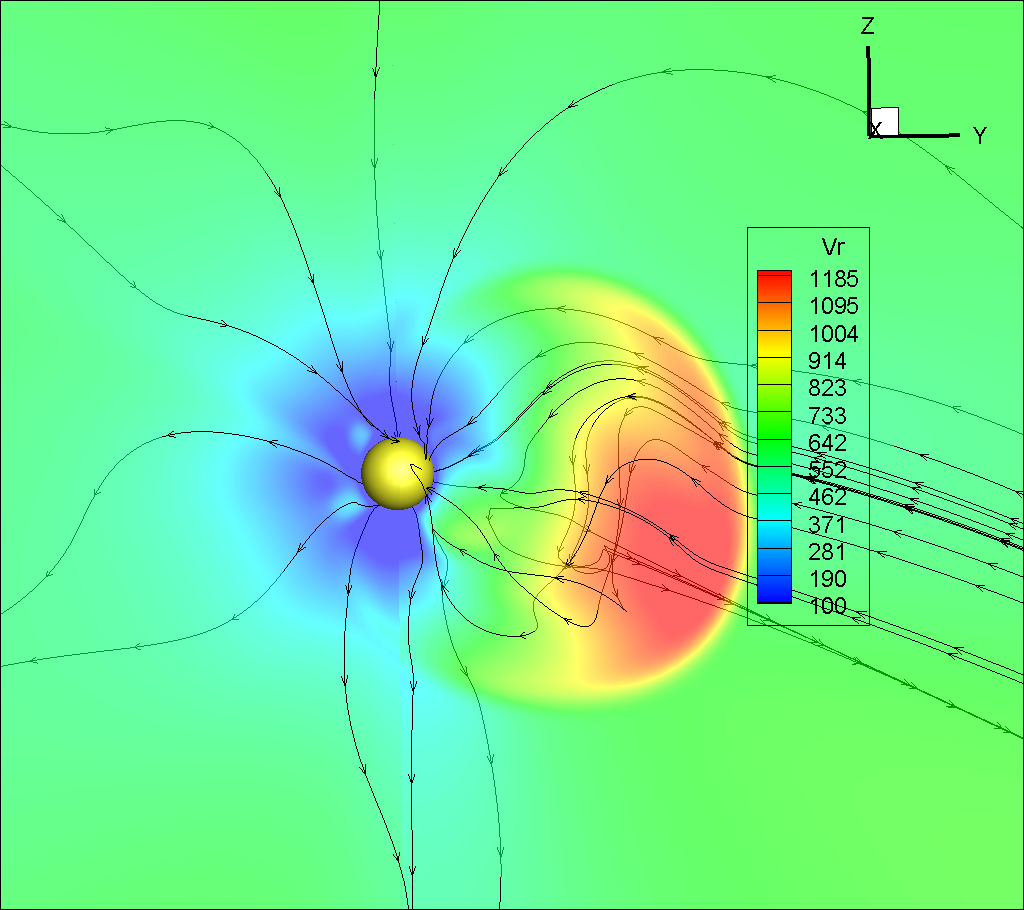} 
\end{tabular}

\caption{(\textit{From left to right}) The colorplot of the radial speed of simulated CME, 17, 34 and 51 minutes after its initial insertion.}
\label{simulation}
\end{figure}

We introduce the modified spheromak into the domain so that the sphere center rests at the inner boundary. This ensures that only one side of the spheromak is introduced into the domain, which resembles a flux rope with two legs (see Fig. \ref{gamma}). The flux rope is introduced so that it replaces the magnetic field in the background corona and the calculated mass of the CME is uniformly distributed over it. We also adjust the parameters $a_1$ and $\gamma$ so that the poloidal and toroidal fluxes introduced into the model match the observations. We still have one free parameter, namely the radius of the spheromak, $r_0$. This parameter can be constrained by the speed of the erupting CME. We find that the speed of eruption depends inversely on the radius of the spheromak, keeping poloidal and toroidal fluxes constant. This can be easily understood considering the source of flux rope eruption is pressure imbalance. If we want to have same flux in a smaller flux rope, we need to increase the magnetic field strength in it. This increases the magnetic pressure inside the flux rope. Therefore, when the solution is evolving in time, smaller flux ropes erupt at greater speeds. \citet{Singh19} reported that the speed of this eruption also depends inversely on the magnetic pressure in the background solar wind in the region in the direction of CME propagation. We keep the initial spheromak in the direction and with the orientation found using GCS method. By varying the value of $r_0$ between 0.7 and 1.0 $R_\odot$, we find that $r_0 = 0.9$ $R_\odot$ results in the speed of the CME as 1200 km/s at 15 $R_\odot$, which is similar to the one calculated from observations. {For this value of $r_0$, the $a_1$ was 0.224 Gauss/$R_\odot^2$ and $\gamma$ was 10.3 for the poloidal and toroidal fluxes to be $1.4\times 10^{22}$ Mx and $2.1\times 10^{21}$ Mx, respectively, which is the average of the observed range}. The parameter $\delta$ was kept as +1, since the flux rope was observed to have positive helicity. In Fig. \ref{white_light}, we show synthetic white-light images to compare the simulated CME shape with the one observed  1.5 hours after eruption. Since we started with a uniform-density flux rope, the CME core is not seen clearly. We do, however, see the bright front and the cavity of this CME. The overall shape of the simulated CME also agrees very well with the observations. This also implies that the initial force distribution in our model is realistic, since a non realistic force distribution would have resulted in considerably different CME shapes compared to observations. Figure \ref{simulation} shows our simulation results for this CME. The CME evolution is shown through snapshots at different time steps. The simulated CME erupts with the proper speed, direction, orientation and magnetic field properties, a key requirement if we want to use MHD modeling for CME predictions. 

\section{Conclusions}\label{Conclusions}
In this study, we modify the spheromak flux rope model so that the poloidal and toroidal fluxes in it can be controlled independently. The motivation for this is the possibility of determination of both these fluxes independently from observations. We show how this model can be used for MHD simulations of CMEs. The pressure imbalance between the flux rope and the surrounding solar wind background leads to its eruption, resembling the characteristics of a CME. In the example shown in this paper, we were able to constrain the speed, direction, orientation, and magnetic properties of the flux rope. We believe that this approach can become a viable option for predicting CMEs, not only in their arrival time, but in their magnetic field properties at 1 AU as well.

TS acknowledges the graduate student support from NASA Earth and Space Science Fellowship. The authors acknowledge the support from the UAH IIDR grant 733033. This work is partly supported by the PSP mission through the UAH-SAO agreement SV4-84017. We also acknowledge NSF PRAC award OAC-1811176 and related computer resources from the Blue Waters sustained-petascale computing project. Supercomputer allocations were also provided on SGI Pleiades by NASA High-End Computing Program award SMD-16-7570 and on Stampede2 by NSF XSEDE project MCA07S033. NG was supported in part by NASA's LWS TR\&T program.

This work utilizes data from \textit{SOHO} which is a project of international cooperation between ESA and NASA. The HMI data have been used courtesy of NASA/\textit{SDO} and HMI science teams. The \textit{STEREO}/SECCHI data used here were produced by an international consortium of the Naval Research Laboratory (USA), Lockheed Martin Solar and Astrophysics Lab (USA), NASA Goddard Space Flight Center (USA), Rutherford Appleton Laboratory (UK), University of Birmingham (UK), Max-Planck-Institute for Solar System Research (Germany), Centre Spatiale de Li\`ege (Belgium), Institut d'Optique Th\'eorique et Appliqu\'ee (France), and Institut d'Astrophysique Spatiale (France). This work uses SOHO CME catalog which is generated and maintained at the CDAW Data Center by NASA and The Catholic University of America in cooperation with the Naval Research Laboratory.

\appendix
\section{Calculating the poloidal and toroidal flux of a flux rope}\label{appendix_A}
\begin{figure}[!htb]
\center

\includegraphics[scale=0.1,angle=0,width=10cm,keepaspectratio]{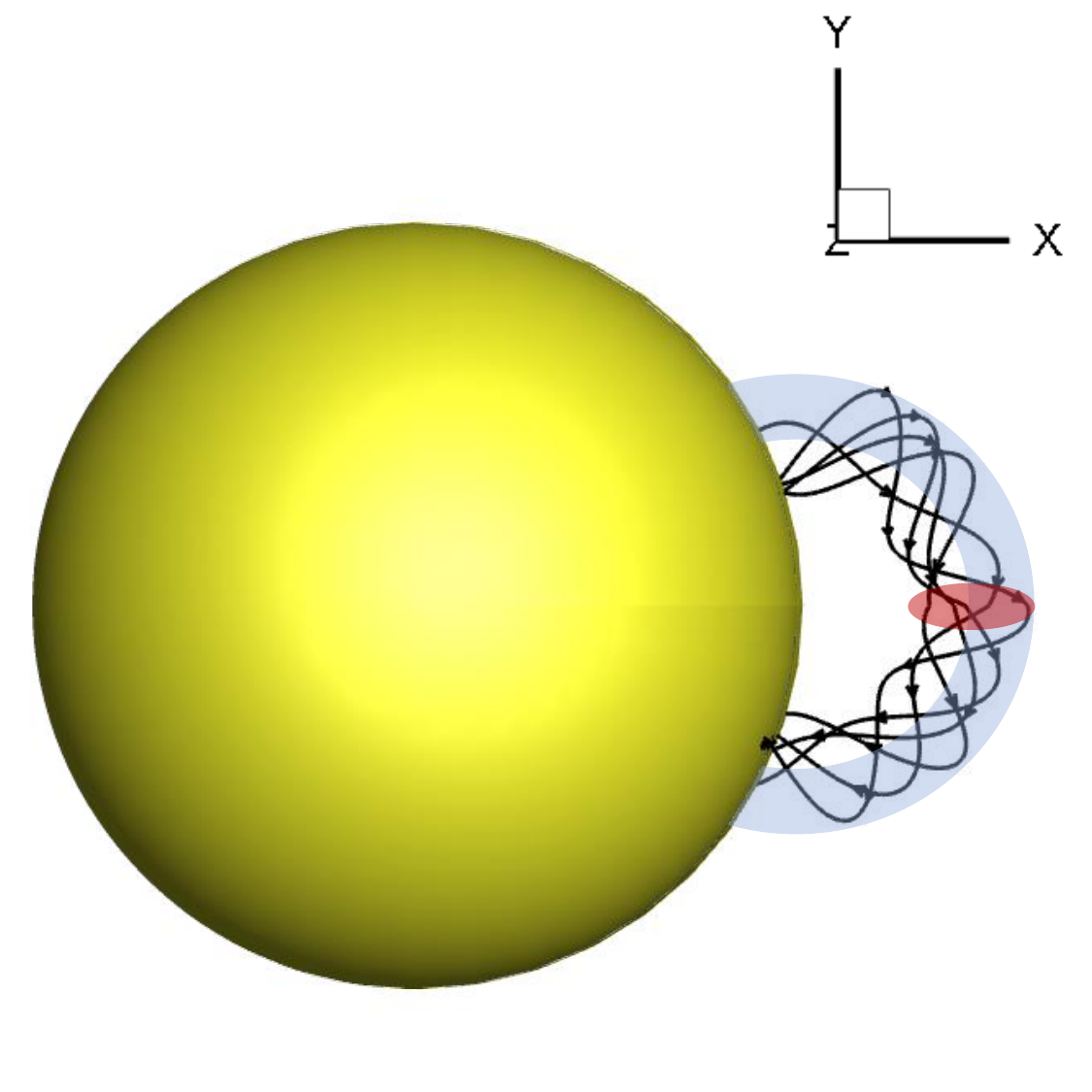}
\caption{A flux rope with its curved axis in the $z=0$ plane is shown anchored to the Sun. The blue shaded region is in the $z=0$ plane and represents the area where $B_z > 0$. The red shaded region is in the $y=0$ plane and represents the cross-section of the flux rope.}
\label{app_a_fig}
\end{figure}

In Fig. \ref{app_a_fig}, we show a flux rope anchored on the Sun, with its curved axis in the $x$-$y$ plane. We shaded the areas used to calculate the magnetic fluxes. The blue region is in the $x$-$y$ plane and represents the area where $B_z > 0$, i.e., field lines are coming out of the plane in this region. The red region represents the cross section of the flux rope and belongs to the $x$-$z$ plane.
The poloidal flux of a flux rope in this configuration is
$$\Phi_p = \int\limits_{Blue\, area} B_z\, dA.$$
The toroidal flux is
$$\Phi_t = \int\limits_{Red\, area} B_y\, dA.$$

\end{document}